\documentclass[12pt]{article}
\usepackage{a4wide}
\usepackage{psfrag}
\usepackage[dvips]{graphicx}
\usepackage{amsfonts}
\usepackage{amstext}
\usepackage{amsmath}
\usepackage{cite}
\renewcommand{\theequation}{\arabic{section}.\arabic{equation}}
\def\a{\alpha}
\def\b{\beta}
\def\i{\infty}
\def\k{\kappa}
\def\m{\mu}
\def\p{\partial}
\def\t{\theta}
\def\T{\Theta}
\def\l{\langle}
\def\r{\rangle}
\def\N{\mathbb{N}}

\begin{document}       
\title{
Statistical Mechanics of an NP-complete Problem:  Subset Sum
}
\author{
\vspace{5mm}
Tomohiro Sasamoto
{\footnote {\tt e-mail: sasamoto@stat.phys.titech.ac.jp 
(Corresponding author.) }},\,
\setcounter{footnote}{2}
Taro Toyoizumi
{\footnote {\tt e-mail: taro@sat.t.u-tokyo.ac.jp}} 
\setcounter{footnote}{3}
and Hidetoshi Nishimori
{\footnote {\tt e-mail: nishi@stat.phys.titech.ac.jp}} \\
{\it $^{*\S}$Department of Physics, Tokyo Institute of Technology,}\\
\vspace{5mm}
{\it Oh-okayama 2-12-1, Meguro-ku, Tokyo 152-8551, Japan}\\
{\it $^{\ddag}$Department of Complexity Science and Engineering,}\\
{\it University of Tokyo,}\\
{\it Hongo 7-3-1, Bunkyo-ku, Tokyo 113-8656, Japan}
}

\date{} 

\maketitle

\begin{abstract}
We study statistical properties of an NP-complete problem,
the subset sum, using the methods and concepts of statistical mechanics.
The problem is a generalization of the number
partitioning problem, which is also an NP-complete problem
and has been studied in the physics literature. 
The asymptotic expressions for the number of solutions 
are obtained. These results applied to the number partitioning 
problem as a special case 
are compared with those which were previously obtained by
a different method. 
We discuss the limit of applicability of the techniques of 
statistical mechanics to the present problem.

\end{abstract}

\section{Introduction}
\label{intro}
The methods and concepts of statistical mechanics have turned
out to be quite useful in the study of problems in computer science
and related fields. 
In particular, the techniques which have originally been developed in 
the spin glass theory have been successfully applied to the 
investigation of the properties of NP-complete problems 
in the theory of computational complexity \cite{GJ}. 
Some of them include 
the travelling salesman \cite{MP}, graph partitioning \cite{FA}, 
$K$-SAT \cite{MZ}, knapsack \cite{I}, vertex cover \cite{WH} 
and other problems \cite{MPV,N}. 
Roughly speaking, NP-complete problems are a class of problems which 
are difficult to solve, in the sense that so far no one has succeeded
in devising (and in fact it is believed to be impossible to design) 
an algorithm to determine in polynomial time 
whether or not there is a solution to given input data. 
NP-complete problems have been extensively studied, 
but still pose many open questions \cite{GJ,Pa}.

The issue of primary interest to computer scientists is to find
an algorithm which efficiently finds a solution to 
given input data, for which purpose statistical mechanics
may not be of direct use because the latter is suitable to reveal
typical properties of many-body systems.
Recently, however, statistical properties of these problems
have been receiving increasing attention since it has gradually been  
recognized that a generally hard problem can sometimes be solved
relatively easily under certain criteria with the assistance of 
statistical mechanics ideas \cite{AI}.

For a wide class of NP-complete problems, the following 
situation happens. A problem has a parameter and,
when the size of the problem becomes large, there
appears a ``critical'' value of the parameter such that below it 
an algorithm can efficiently find a solution (easy region) but 
above it the same algorithm no longer works effectively (hard region).
This happens because the definition of NP-completeness is based on the
{\it worst} case analysis. A problem can be classified as a difficult one
if there are only a few difficult instances.
The sudden change of the statistical properties of 
a problem is in many respects similar to a phase transition, 
a concept from statistical mechanics.
In fact, the methods for studying phase transitions have turned out 
to be powerful tools to understand the properties of the 
above-mentioned phenomena. 
These observations suggest that the 
{\it typical} case study will play increasingly important roles
in computer science and accordingly the methods from statistical 
mechanics will provide useful tools. 

In typical case studies, one usually considers a randomized version of a 
problem. 
In other words, our main interest is in the properties of 
the problem averaged over possible realizations of input data.
The randomized problems share many features with spin glass systems
and have been often studied using the techniques of the spin glass theory.
In particular the replica method has allowed us to analyze the 
problems, many of which would have been impossible to deal with without it.
Nonetheless the resulting saddle point analysis, known as the problem 
of replica-symmetry breaking,  is often so hard that
it is usually difficult to get complete understanding of the problem.
Hence, to gain more insights, it is important to study problems 
which are solvable without using replicas.

The number partitioning problem seems to be
an ideal example from this point of view \cite{Fu,FF,mPRL98,mP00}.  
Suppose that one is given a set of positive integers 
$\mathcal{A}=\{a_1,a_2,\ldots,a_N\}$ and asked to divide 
this into two subsets with the same value of the sums. 
In other words, one tries to find a subset 
$\mathcal{A}'\subset\mathcal{A}$ which minimizes the partition difference
\begin{equation}
  \left| \sum_{\mathcal{A}'} a_j
        -\sum_{\mathcal{A}\backslash \mathcal{A}'} a_j\right|.
\end{equation}
A subset $\mathcal{A}'$ with zero partition difference is called a 
perfect partition, whereas a subset with a positive partition difference
is termed an imperfect partition.
It has been argued that 
this problem shows a sharp change of states, reminiscent of a phase 
transition, between easy and hard regions \cite{GW,mPRL98,FF,mP00}.
In addition, the problem has a lot of practical 
applications such as multiprocessor scheduling and minimization of 
VLSI circuit size.

The analysis in \cite{mPRL98,mP00} starts from taking the partition difference
to be the Hamiltonian. Then a perfect partition corresponds
to a ground state of the Hamiltonian and an imperfect partition 
to a configuration with positive energy.
By applying the statistical mechanics methods and a saddle point  
approximation in the large-$N$ limit, several results have been 
obtained without using replicas.
The phase transition behaviour of the problem, found numerically
\cite{GW}, was understandable through those results.
But the expressions obtained in his analysis show peculiar
high temperature behaviours as will be shown below. 
In particular, the partition 
function does not give the correct entropy in the limit of high
temperature. Hence his results are not expected to give reliable
predictions for imperfect partitions.

The main purpose of this paper is to propose an alternative approach 
to the number partitioning problem applicable to
imperfect partitions as well.
We study a generalized version of the number 
partitioning problem: the subset sum \cite{GJ}.
By using some basic concepts and methods of statistical mechanics,
the asymptotic expressions of the number of solutions are obtained.
Our results specialized to the number partitioning problem
are compared with the previously obtained predictions.
It is shown that our results are applicable to the cases where the 
predictions of the previous analysis do not agree with 
an exactly solvable example.
Our discussions are mainly restricted to the easy region although
the hard region could also be considered by similar arguments using the
ideas in \cite{mPRL00}.

The rest of the paper is organized as follows.
In the next section, we introduce the subset sum and
reformulate it in terms of a Hamiltonian. 
By using the canonical ensemble, 
the asymptotic number of solutions is estimated in section 3.
Based on the results, we discuss a crossover between 
easy and hard regions of the subset sum in section 4.
In section 5, the analysis is generalized to the case with
constraint. 
In section 6, we apply the results to the number partitioning problem
and compare the results with those in \cite{mPRL98}.
Conclusion is given in the last section.

\setcounter{equation}{0}
\section{Subset Sum}
Let us denote $\N_+=\{1,2,\cdots\}$, the set of positive integers.
The subset sum is an example of NP-complete problems in which one
asks, for a given set of $\mathcal{A}=\{a_1,a_2,\cdots,a_N\}$ 
with $a_j \in \N_+$ \,$(j=1,2,\cdots,N)$  and $E \in \N_+$, 
whether or not there exists a subset $\mathcal{A}'\subset A$ 
such that the sum of the elements of $\mathcal{A}'$ is $E$ \cite{GJ}.
To formulate the problem, we introduce a Hamiltonian (or energy)
\begin{equation}
  \label{hami}
  H=\sum_{j=1}^N a_j n_j,
\end{equation}
where $n_j\in\{0,1\}$ \,$(j=1,2,\cdots,N)$, and
the subset sum is equivalent to asking whether or not 
there exists a configuration $\{n_1,n_2,\cdots,n_N\}$ such that $H=E$.
A configuration which satisfies $H=E$ is called a solution in the following.

There are several versions of the problem.
The original one is the decision problem;
one only asks whether there exists a solution or not.
Once one learns that the answer to the decision problem is yes, however, 
it would be quite natural next to ask how many solutions there are. 
This is called the counting (or enumeration) version of the problem.
On the other hand, if there is no solution, one 
might try to find the best possible configuration which 
minimizes the energy difference from the given $E$.
This is the optimization version of the problem.
Of course these versions are closely related to each other.
In the following treatments, we focus on the counting version 
of the problem, for which statistical mechanics 
provide powerful analytical tools.
The number of solutions for a given energy $E$ will be denoted by $W(E)$.

\setcounter{equation}{0}
\section{Statistical Mechanical Analysis of Subset Sum}
Evaluation of the exact value of $W(E)$ for given 
$\mathcal{A}$ and $E$ is still a question of complicated combinatorics
and is very hard. In particular, fixing the value of $E$ is a
very strong constraint which renders the counting almost intractable.
In the terminology of statistical mechanics, 
considering the problem with a fixed value of $E$ 
corresponds to working in the microcanonical ensemble.
For many purposes in practice, however, one is interested in 
the asymptotic behaviours for a large $N$ and is satisfied with 
approximate expressions of $W(E)$; an exact expression of $W(E)$ 
is unnecessary and can even obscure the essential aspects of the problem. 
The experience in the study of statistical mechanics tells us that, 
in order to know the asymptotic behaviour of $W(E)$, it is 
much easier to work in the canonical ensemble.  
This is a superposition of the microcanonical ensembles
for all possible values of the energy 
with the Boltzmann factor $e^{-\b E}$,
where $\b$ is the inverse temperature. 
In this section, the set $\mathcal{A}$ is still fixed;
statistics over many $\mathcal{A}$ is not considered.

Simplicity of the analysis of the subset sum compared with 
other problems 
stems from a compact expression for the partition function.
For a given $\mathcal{A}$, the partition function $Z$ 
is simply given by
\begin{align}
  \label{Z}
  Z &=\sum_{\{n_j\}} e^{-\b H}
     =\sum_{n_1=0,1}\sum_{n_2=0,1}\cdots\sum_{n_N=0,1} e^{-\b H} \notag\\
    &=(1+e^{-\b a_1}) (1+e^{-\b a_2}) \cdots (1+e^{-\b a_N}).
\end{align}
 From this one can calculate the average values of various physical quantities.
The average here means the thermal average and is denoted by
$\l \cdots \r$.
The average energy $\l E \r$ for a given value of $\b$ is given by 
\begin{equation}
  \l E \r = -\frac{\p}{\p \b} \log Z
          = \sum_{j=1}^N \frac{a_j}{1+e^{\b a_j}}.
\end{equation}
Note that the value of $\l E\r$ can be controlled  by changing $\b$.
As $\b$ is increased from $-\i$ to $\i$, the average energy 
$\l E\r$ decreases from $\sum_{j=1}^N a_j$ to $0$.
In usual statistical mechanics,
the temperature and hence $\b$ should be positive.
For our present problem, however, the temperature 
is introduced only as a parameter to control the average energy. 
A negative value of $\b$ is also allowed in our problem.
The fluctuation of the energy is similarly calculated as
\begin{equation}
  \label{Eflu}
  \l (E - \l E \r)^2 \r 
  = 
  \frac{\p^2}{\p \b^2} \log Z
  =
  \sum_{j=1}^N \frac{a_j^2}{(1+e^{\b a_j})(1+e^{-\b a_j})}.
\end{equation}

Here we go back to (\ref{Z}) and 
observe that $W(E)$, the number of solutions to the condition $H=E$, 
appears as the coefficient of the $E$th power of $q$ ($=e^{-\b}$) in $Z$;
the expansion of $Z$ in terms of $q$ gives 
\begin{equation}
  \label{Zex}
  Z = \sum_{E=0}^{E_{\text{max}}} W(E) \,q^E
\end{equation}
with $E_{\text{max}}=a_1+a_2+\cdots + a_N$.
Moreover, since $Z$ is a polynomial in $q$, (\ref{Zex}) can be
inverted easily: $W(E)$ has 
an integral representation
\begin{equation}
  \label{Wcont}
  W(E) = \int_{C}\frac{dq}{2\pi i} Z q^{-E-1}
\end{equation}
with $C$ being a contour enclosing the origin anticlockwise 
on the complex $q$ plane.
It is important to notice that the contour $C$ in (\ref{Wcont})
can be deformed arbitrarily as far as it encloses the origin anticlockwise.

Now we consider the asymptotics of $W(E)$ as $N\to\i$.
One should specify how this limit is taken since 
changing $N$ also implies changing $\mathcal{A}$ simultaneously.
To avoid this difficulty, let us suppose for the moment 
that one first has an infinite set
$\mathcal{A}_\i=\{a_1,a_2,\cdots\}$, each element of which 
is taken from a finite set of $\{1,2,\cdots,L\}$, with $L\in\N_+$.
Then the set $\mathcal{A}$ can be regarded as a collection of the 
first $N$ elements of $\mathcal{A}_\i$. 
The limit $N\to\i$ is defined without ambiguities in this way.

Since $a_j$ satisfies $1\leq a_j\leq L$ for all $j$, 
a simple estimation of (\ref{Eflu}) shows that the fluctuation of
the energy is of order $N$ when $\b$ is finite.
Hence the fluctuation of the energy per system size $N$, i.e., 
$\l (E/N-\l E \r/N)^2\r$, tends to zero as $N\to\i$.
Let $\b_0$ be the value of $\b$ such that the average energy 
$\l E\r$ is equal to $E$.
Then one takes the contour $C$ in (\ref{Wcont}) to be a circle with 
radius $e^{-\b_0}$ and uses the phase variable $\t$ defined by 
$q=e^{-\b_0+i\t}$ to find
\begin{equation}
  \label{Wtheta}
  W(E) =\frac{1}{2\pi}\int_{-\pi}^{\pi} d\t e^{\log Z+\b_0 E -iE\t}.
\end{equation}
It is not difficult to verify that the saddle point of this integrand 
is at $\t=0$. (When the g.c.d. of $\mathcal{A}$ is not one, 
there appear other saddle points with the same order of contributions.
But, when $N$ is large enough, it is almost sure that the g.c.d. of 
$\mathcal{A}$ is one, which we assume in what follows.)

The quantities in the exponent on the right hand side of (\ref{Wtheta})
are all of order $N$. Therefore we can use the method of steepest
descent to evaluate the asymptotic behaviour of the integral.
Expanding $\log Z$ around $\b=\b_0$ to second order leads to
\begin{equation}
  W(E) = e^{\log Z|_{\b=\b_0} + \b_0 E}\cdot
         \frac{1}{2\pi}\int_{-\pi}^{\pi}d\t
         \exp\left(-\frac{\t^2}{2}\frac{\p^2}{\p \b^2} \log Z|_{\b=\b_0}\right).
\end{equation}
Since the second derivative of $\log Z$ is the
fluctuation of the energy (\ref{Eflu}) and is of order $N$, 
the integration range may be extended to $\pm\i$. The result is
\begin{equation}
  W(E)  
  \approx
  \dfrac{\exp\left[ \log Z|_{\b=\b_0} + \b_0 E \right]}
        {\sqrt{2\pi \frac{\p^2}{\p \b^2} \log Z|_{\b=\b_0}}}.
\end{equation}
Here and in the following the symbol $\approx$ means that the ratio of 
the right and left hand sides tends to unity as $N\to\i$.
After rewriting $\b_0$ back to $\b$, we finally obtain the asymptotic
expression of $W(E)$ for a given value of $E$ via a common parameter $\b$ as 
\begin{align}
  \label{W_b}
  W(E)
  &\approx 
  \dfrac{\exp\left[ \sum_{j=1}^N 
                  \log(1+e^{-\b a_j})+\b \sum_{j=1}^N a_j/(1+e^{\b a_j}) 
            \right]}
     {\sqrt{2\pi \sum_{j=1}^N a_j^2/(1+e^{\b a_j})(1+e^{-\b a_j})}} ,
  \\
  \label{E_b}
  E &= \sum_{j=1}^N \frac{a_j}{1+e^{\b a_j}}. 
\end{align}
A numerical check of (\ref{W_b}) and (\ref{E_b}) is shown
in Fig. 1.
As far as one sees on this scale, the agreement of our predictions 
and simulational data is satisfactory for the entire range of the energy.

One may notice that the obtained expressions, (\ref{W_b}) and
(\ref{E_b}), do not depend on $L$. 
This is plausible since the number of solutions depends only
on the elements of $\mathcal{A}$, not directly on the set from which
elements of $\mathcal{A}$ have been taken. 
One should remember in this relation that 
the validity of the saddle point analysis depends on $L$. 
The expressions, (\ref{W_b}) and (\ref{E_b}), 
become better approximations as $N\to\i$ for a fixed value of $L$. 
It would not be surprising if the expressions do not agree very well
with simulational
data when $L$ is sufficiently large so that $W(E)$ is of $O(1)$.
In particular one should not use (\ref{W_b}) in the parameter
region which gives $W(E)<1$ as will be discussed below.

\setcounter{equation}{0}
\section{Easy/Hard Regions}
Numerical simulations in \cite{GW} suggest that there exist 
easy and hard regions for a randomized version of the subset sum,
in which one considers statistics over many samples of $\mathcal{A}$
with each $a_j$ drawn from $\{1,2,\ldots,L\}$ uniformly.
In simulations, one checks if there exists a solution
for many samples of $\mathcal{A}$ with given $N,L$ and $E$. 
Then for each $N$ and $E$, one plots the probability that there is 
at least one solution as a function of $\k=\log_2 L/N$.
Then it is observed that the probability decreases fairly sharply 
from 1 to 0 as $\k$ increases from zero to $\i$. 
As $N$ becomes larger, the decrease of the probability occurs 
in a narrower range of $\k$. 
In fact, the system appears to have a sharp transition at a critical value
$\k_c$ in the limit of $N,L\to\i$ \cite{GW}.
In this section, we estimate the critical value $\k_c$ of the randomized 
subset sum by using the results of the last section.

The analysis in the last section gives us the asymptotic 
formula for a fixed $\mathcal{A}$. 
To apply the results to the randomized version of the problem,
one has to notice that, as $N$ becomes large,  
the sample dependence of (\ref{W_b}) and (\ref{E_b}) is suppressed 
increasingly.
In fact, in the limit $N\to\i$ with $L$ fixed, 
there would be no sample dependence 
so that the average properties of these quantities coincide with those
of a typical sample.  
To see this, let us define the density of $y_j=a_j/L$ $(1\leq j\leq N)$
to be $\rho_N(y)=\frac{1}{N}\sum_{j=1}^N \delta(y-y_j)$.
Since we draw the $a_j$ uniformly, we have
$\displaystyle\lim_{N\to\i}\rho_N(y) = \rho(y)$ where 
$\rho(y)=1$ for $0\leq y\leq 1$ and $\rho(y)=0$ otherwise. 
In addition, in this limit, 
summations in (\ref{W_b}) and (\ref{E_b}) are replaced by integrals,
resulting in 
\begin{gather}
  W(E)  \approx 
         \dfrac{\exp\left[ N \int_0^1 dy 
                   \left\{\log(1+e^{-\a y})+\a y/(1+e^{\a y})\right\}
            \right]}
         {\sqrt{2\pi N L^2 \int_0^1 dy y^2/(1+e^{\a y})(1+e^{-\a y})}},  \\    
  \label{E_a}
  x=\frac{E}{N\cdot L} \approx   \int_0^1 dy \frac{y}{1+e^{\a y}},
\end{gather}
where we have introduced a scaled inverse temperature $\a = \b/L$.
The parameter $\a$ controls $x$, the energy divided by $N\cdot L$;
as $\a$ is increased from $-\i$ to $\i$, $x$ decreases from $1/2$ to $0$.
The validity of these expressions is determined only by the 
values of $N$ and $L$. 
For a given $N$, they are valid for sufficiently small $L$.
Even though $L$ changes, however, these expressions 
are expected to be good approximations as long as 
$N$ is relatively large or when $\k$ is fairly smaller than $\k_c$. 
On the other hand, the reliability of these expressions 
is unclear for $\k \gg \k_c$.
In fact, there is evidence that the average minimal cost
is not self-averaging in this region \cite{FF}.
The value of $\k$ below which the above formulas are valid increases as 
$N$ and $L$ increase, and finally it reaches $\k_c$ in the limit $N,L\to\i$. 
It is important to notice that 
the value of $\k_c$ can be determined by the condition $W(E)=1$
in the limit $N,L\to\i$ because $W(E)$ is the expectation value of
the number of configurations. We therefore find
\begin{equation}
  \k_c = \frac{1}{\log 2}
         \int_0^1 dy \left[\log(1+e^{-\a y})+\frac{\a y}{1+e^{\a y}}\right]
\end{equation}
with $\a$ determined by (\ref{E_a}) for a given value of $x$.
For $\k<\k_c$, exponentially many solutions are expected to exist and 
one of them can be found fairly easily.
On the other hand, for $\k>\k_c$, there is practically no solution and 
hence it is virtually impossible to find one. 
The easy/hard regions of the randomized subset sum are shown in Fig. 2.

\setcounter{equation}{0}
\section{Constrained Case}
In some applications, one might encounter a situation where 
the number of $a_j$'s is given.
In this section, our previous analysis is generalized to 
the {\it constrained} case
where the number of chosen $a_j$'s is fixed to $M$.
Instead of considering directly the system with 
constraint, we again take a superposition of the problems with 
various values of $M$.
In the language of statistical mechanics, we work in 
the grand canonical ensemble.
Let us define a Hamiltonian
\begin{equation}
  H_{\text{c}} = \sum_{j=1}^N a_j n_j -\frac{\m}{\b}\sum_{j=1}^N n_j.
\end{equation}
The first term is nothing but the Hamiltonian (\ref{hami}) 
for the unconstrained subset sum.
The second term is introduced to control the number of $a_j$'s
by changing the parameter $\m$, the chemical potential.
The grand partition function is evaluated as
\begin{align}
  \T &= \sum_{\{n_j\}} e^{-\b H_{\text{c}}} \notag\\
     &= (1+e^{\m}e^{-\b a_1})(1+e^{\m}e^{-\b a_2})
        \cdots(1+e^{\m}e^{-\b a_N}) \notag\\ 
     &= \sum_{M=0}^N\sum_{E=0}^{E_{\text{max}}}
        W(M,E) e^{\m M} e^{-\b E},
\end{align}
with $E_{\text{max}}=a_1+a_2+\cdots +a_N$ as before.
Here $W(M,E)$ is the number of configurations which satisfy
$\sum_{j=1}^N a_j n_j=E$ and $\sum_{j=1}^N n_j=M$ 
simultaneously.

For given values of $\m$ and $\b$,
the average number, energy and second moments of these quantities are
expressed as
\begin{align}
  \label{gM}
  \l M \r 
  &= 
  \frac{\p}{\p \m} \log \T
  =
  \sum_{j=1}^N \frac{1}{1+e^{\b a_j-\m}} , \\
  \label{gE}
  \l E \r 
  &= 
  -\frac{\p}{\p \b} \log \T
  =
  \sum_{j=1}^N \frac{a_j}{1+e^{\b a_j-\m}} , \\
  \l (M-\l M \r)^2 \r 
  &= 
  \frac{\p^2}{\p \m^2} \log \T
  =
  \sum_{j=1}^N \frac{1}{(1+e^{\b a_j-\m})(1+e^{-\b a_j+\m})} ,\\
  \l (M-\l M \r)(E-\l E \r) \r 
  &= 
  -\frac{\p^2}{\p \b \p\m} \log \T
  =
  \sum_{j=1}^N \frac{a_j}{(1+e^{\b a_j-\m})(1+e^{-\b a_j+\m})} ,\\
  \l (E-\l E \r)^2 \r 
  &= 
  \frac{\p^2}{\p \b^2} \log \T
  =
  \sum_{j=1}^N \frac{a_j^2}{(1+e^{\b a_j-\m})(1+e^{-\b a_j+\m})}. 
\end{align}
Similarly to the unconstrained case, one can show that the fluctuations of the 
number and energy divided by the system size vanish as $N\to\i$ so
that one can apply the saddle point method.
The resulting asymptotic expression for $W(M,E)$ reads
\begin{equation}
  W(M,E) 
  \approx
  \frac{\exp[\log\T+\b E- \m M]}
       {2\pi\sqrt{D}}
\end{equation}
with $D$ being
\begin{equation}
  D=
  \begin{vmatrix}
    \frac{\p^2}{\p\m^2}  \log \T & -\frac{\p^2}{\p \b\p\m}\log \T\\
   -\frac{\p^2}{\p\b\p\m}\log \T &  \frac{\p^2}{\p \b^2} \log \T .
  \end{vmatrix}
\end{equation}
The values of $M$ and $E$ are given by (\ref{gM}) and (\ref{gE}),
respectively.
Using these expressions, one can discuss the easy/hard regions
of the constrained subset sum. 
The analysis is almost the same as that in the last section
and is omitted here.

\setcounter{equation}{0}
\section{Number Partitioning Problem}
As already mentioned in the introduction,
the subset sum is regarded as a generalization of the 
number partitioning problem. 
In this section, we apply our previous discussions to 
the number partitioning problem.
Our results are compared with those in \cite{mPRL98,mP00},
which are briefly reviewed in Appendix A with some remarks.

Let us first establish an explicit relationship between the 
subset sum and the number partitioning problem.
If one introduces the spin variables by 
\begin{equation}
  s_j = 2 n_j-1,
\end{equation}
it is not difficult to see
\begin{equation}
  \label{npp_ss}
  \tilde{H} := \left| 2 H -\sum_{j=1}^N a_j \right| 
             = \left| \sum_{j=1}^N a_j s_j  \right|.
\end{equation}
This is exactly the Hamiltonian of the number partitioning 
problem studied in \cite{mPRL98,mP00}.

The number of solutions of $\tilde{H}=\tilde{E}$, which we denote 
by $\tilde{W}(\tilde{E})$, is related to $W(E)$ by
\begin{align}
  \tilde{W}(\tilde{E}) 
  =
  \begin{cases}
   W(\frac12 \sum_{j=1}^N a_j)  
   &(\tilde{E}=0)\\
    W( \frac12 \tilde{E}+\frac12 \sum_{j=1}^N a_j)
   +W(-\frac12 \tilde{E}+\frac12\sum_{j=1}^N a_j)
   &(\tilde{E}>0)
  \end{cases}.
\end{align}
Of special interest is the case of $\tilde{E}=0$, a solution of 
which is called a perfect solution in the number 
partitioning problem.
In the subset sum, this corresponds to the energy
$E = \frac12\sum_{j=1}^N a_j$. Clearly there is no perfect solutions
if $\sum_{j=1}^N a_j$ is odd; we assume $\sum_{j=1}^N a_j$ is even 
in the following.
In terms of $\b$, considering perfect solutions corresponds to $\b= 0$
from (\ref{E_b}).   
Setting $\b=0$ in (\ref{W_b}) leads to
\begin{align}
  \label{W_b0}
  \tilde{W}(0) 
  \approx 
  \dfrac{2^N}{\sqrt{\frac{\pi}2  \sum_{j=1}^N a_j^2}}, 
\end{align}
which is expected to be the number of perfect solutions to 
the number partitioning problem. 
In fact (\ref{W_b0}) agrees with the previously obtained
result in \cite{mPRL98,mP00}.
Hence, as far as the number of perfect solutions for the number 
partitioning problem is concerned, our method gives exactly the same
answer as in \cite{mPRL98,mP00}.

The difference between our formula and that in \cite{mPRL98,mP00}
becomes manifest for a finite value of $\tilde{E}$.
We demonstrate this by considering the number partitioning problem for 
a special case where $a_1=a_2=\cdots=a_N=1$ with $N$ even.
In this case, the Hamiltonian reads $\tilde{H}=|\sum_{j=1}^N s_j|$, and 
it is possible to write down the partition function 
$\tilde{Z}=\sum_{\{s_j\}}e^{-\b \tilde{H}}$ explicitly:
\begin{align}
  \tilde{Z} 
  &=\sum_{j=1}^N \binom{N}{j} e^{-\b|N-2j|} \notag\\
  &=\binom{N}{\frac{N}2} +2 \sum_{j=1}^{N/2} \binom{N}{\frac{N}2+j} e^{-2\b j}.
\end{align}
This formula indicates that there are solutions for even $\tilde{E}$ and 
that $\tilde{W}(\tilde{E})$ is 
\begin{equation}
    \label{a1}
\tilde{W}(\tilde{E})
=
\begin{cases}
    \displaystyle\binom{N}{N/2}   \approx  \dfrac{2^N}{\sqrt{\frac{\pi}2 N}}    
                            & (\tilde{E}=0)  \\
  \displaystyle 2 \binom{N}{N/2+j} \approx  
  \dfrac{2\exp N \left[-(\frac12-\frac{j}N)\log(\frac12-\frac{j}N)
                            -(\frac12+\frac{j}N)\log(\frac12+\frac{j}N)\right]}
              {\sqrt{2\pi N(\frac12+\frac{j}N)(\frac12-\frac{j}N)}}   
                            & (\tilde{E}=2j)
  \end{cases},
\end{equation} 
where the asymptotics are also indicated.
For the present case with $a_j=1$ ($1\leq j\leq N$), 
(\ref{E_b}) is simply reverted as $\b=\log(N/E-1)$. 
Then, using Stirling's formula, one can confirm 
that our formula (\ref{W_b}) gives correct asymptotics in (\ref{a1})
for the entire range of energy.
By contrast the partition function in \cite{mPRL98,mP00}, which 
we denote by $\tilde{Z}'$, for the 
present case can be written as
\begin{equation}
  \label{MerZ}
  \tilde{Z}'
  =
  \frac{2^N}{\sqrt{\frac{\pi}2 N}}\left(1+2\sum_{j=1}^\i e^{-2\b j}\right).
\end{equation}
This indicates that there are solutions for even $\tilde{E}$ and 
that $\tilde{W}(\tilde{E})$ is asymptotically  
$2^N/\sqrt{\frac{\pi}2 N}$ for $\tilde{E}=0$ and 
$2^{N+1}/\sqrt{\frac{\pi}2 N}$ for $\tilde{E}=2j$. 
As is clear from (\ref{a1}),
the correct asymptotics is predicted only for $\tilde{E}=0$.
One may notice that the arguments in \cite{mPRL98,mP00} 
are somewhat different from ours.
There, the energy $\tilde{E}$ and the entropy $\tilde{S}$ 
are calculated from the 
partition function (\ref{MerZ}), following the usual prescriptions 
of statistical mechanics. It is assumed that $\exp({\tilde{S}})$ gives 
the number of solutions. The obtained expressions again do not give the 
correct asymptotics of $\tilde{W}(\tilde{E})$ when $\tilde{E}>0$.
Our conclusion is that the results of \cite{mPRL98,mP00} 
give the correct asymptotic value for $\tilde{E}=0$ 
but not for $\tilde{E}>0$. 

The reason for this difficulty is traced back to the application of
statistical mechanics techniques to the system for which the number of
solutions of $\tilde{H}=\tilde{E}$ decreases as $\tilde{E}$
increases. It can be seen from Fig. 1 that the number partitioning
problem is indeed an example with this anomalous property if one notes
that $\tilde{E}=0$ corresponds to the peak of the curve.
The problem is that, for such systems, usual prescriptions 
of the canonical ensemble do not work.
For normal physical systems, the number of states $W(E)$ increases as 
a function of the energy $E$. 
The number of states multiplied by the Boltzmann factor 
$W(E)e^{-\b E}$ takes a maximum at some value of $E$. 
The peak around this point becomes drastically sharp 
as the system size increases. 
Then the equivalence of microcanonical and canonical 
ensembles holds so that we can study the 
thermodynamic behaviours of the system in either ensemble.
For systems with decreasing 
$W(E)$, however, $W(E)e^{-\b E}$ is a monotone decreasing function. 
The fluctuation of the energy does not tend to zero
even when the system size increases indefinitely, and consequently 
one can not control the energy by changing the temperature. 
As a result, the equivalence of ensembles does not hold.
The exponential of the entropy calculated in the canonical ensemble
and the coefficient of the expansion of $Z$ in powers of $e^{-\b E}$
do not agree even in the thermodynamic limit; 
in addition, neither of these quantities
give the correct asymptotics of the number of configurations $W(E)$.
This problem may be overcome by considering a negative
temperature as we did for the subset sum, but a direct
analysis of the number partitioning problem described by
(\ref{npp_ss}) would then be much more difficult.

Before closing this section, we discuss the randomized number 
partitioning problem with a constraint in which one asks whether or not
there exists a perfect solution with $\sum_{j=1}^N s_j$  fixed.
In the language of subset sum, this corresponds to fixing $M$ since
$\sum_{j=1}^N s_j=\sum_{j=1}^N (2 n_j-1)=2 M -N$.
In \cite{FF,mP00}, it has been found that there is a phase
transition in the limit $M, N\to\i$ with $m=2M/N-1$ fixed.
We can reproduce this phenomenon from the results of section 5.
In the limit $N,L$ and $M\to\i$, summations in (\ref{gM}) and (\ref{gE}) 
are replaced by integrals. They are written as, for a uniform distribution,
\begin{align}
  \label{Cm}
  \frac12(1+m) 
       &= \int_0^1 d y \frac{1}{1+e^{\a y-\m}}
        =1-\frac{1}{\a}\log\frac{e^{\a}+e^{\m}}{1+e^{\m}}, \\
  \label{Cx}
  x    &= \int_0^1 d y \frac{y}{1+e^{\a y-\m}}, 
\end{align}
where $\a=\b/L$ and $x=E/N\cdot L$ as before.
Since (\ref{Cm}) is easily reverted as
\begin{equation}
  \m = \log \frac{e^{\a}-e^{\a(1-m)/2}}{e^{\a(1-m)/2}-1},
\end{equation}
one can regard $x$ as a function of $\a$ and $m$.
Then, for a fixed $m$ ($-1\leq m\leq 1$), one sees that 
$x$ decreases from $(1+m)(3-m)/8$ to $(1+m)^2/8$
as $\a$ is increased from $-\i$ to $\i$. 
There are few or no configurations 
for energy corresponding to $x$ outside this range. 
If we note that the perfect solution corresponds to $x=1/4$,
we find that there are extensive number of perfect 
solutions when $|m|<\sqrt{2}-1$ and 
there is practically no perfect solution when
$|m|>\sqrt{2}-1$. Hence we conclude that there is a phase transition
at $m_c=\sqrt{2}-1$, in agreement with the previous analysis 
\cite{FF,mP00}.

\setcounter{equation}{0}
\section{Conclusion}
We have studied the statistical properties of the subset sum, which is a
generalization of the number partitioning problem.
The basic ideas and methods of statistical mechanics enabled us to study the
asymptotic behaviour of the number of solutions for a given set of input data.
The expressions (\ref{W_b}) and (\ref{E_b}) represent the main results 
of this paper. The agreement of the predictions with simulational 
data have been found satisfactory. Our results have been compared 
with those which were previously obtained by a different method.
They agreed with each other for the number of perfect solutions 
of the number partitioning problem. On the other hand, in the 
case of the subset sum, only our analysis gave the correct asymptotics
over the entire range of energy.
The reason why the validity of the results in \cite{mPRL98,mP00} 
is restricted to perfect solutions has been argued to be that 
the entropy calculated in canonical ensemble 
does not necessarily give the logarithm of the number 
of configurations for systems with a decreasing number of states
as the energy increases. In such anomalous systems, one should
be extremely careful in using the equivalence between microcanonical,
canonical, and grand canonical ensembles.

\section*{Acknowledgment}
The work of HN was supported by the Grant-in-Aid for 
Scientific Research by the Ministry of 
Education, Culture, Sports, Science and Technology,
the Sumitomo Foundation and the Anglo-Japanese Collaboration
Programme between the Japan Society for the Promotion of 
Science and the Royal Society.
TS acknowledges the Grant-in-Aid for Encouragement of Young Scientists 
from the Ministry of Education, Culture, Sports, Science and Technology.


\setcounter{equation}{0}
\setcounter{section}{0}
\renewcommand{\thesection}{Appendix \Alph{section}}
\renewcommand{\theequation}{\Alph{section}.\arabic{equation}}
\section{Some Remarks on the Analysis in \cite{mPRL98,mP00}}
In this Appendix, we briefly review the results in \cite{mPRL98,mP00}
and give a few remarks on the analysis.
Our notations are slightly different from the original ones
for consistency with the main text.
The g.c.d. of $\mathcal{A}$ is assumed to be 
unity in the following.
After some manipulations,
the partition function $\tilde{Z}$
for the Hamiltonian (\ref{npp_ss}) is rewritten as
\begin{equation}
  \label{MZ}
  \tilde{Z} = \sum_{\{s_j\}} e^{-\b \tilde{H}}
            = 2^N \int_{-\pi/2}^{\pi/2}\frac{dy}{\pi} e^{N G(y)},
\end{equation}
where $\beta (\ge 0)$ is the inverse temperature and
\begin{equation}
  G(y) = \frac{1}{N} \sum_{j=1}^N\log\cos (\beta a_j \tan y). \label{3-G}
\end{equation}
Then it has been argued that, for a large $N$, the integral in (\ref{MZ})
can be evaluated using the Laplace method.
There exist an infinite number of points which give the maximum
value of $\text{Re}\{G(y)\}$.
The main contributions are expected to come from the points
\begin{equation}
  y_k=\arctan \left(\frac{\pi}{\beta} k\right), \quad k=0,\pm 1,\pm 2,\dots .
\label{3-sp}
\end{equation}
It is not difficult to confirm $\text{Re}\{G(y)\}\leq 0$ for
general $y$ and $\text{Re}\{G(y_k)\}=G'(y_k)=0$ and $G''(y_k)<0$, 
so that $y_k$'s of (\ref{3-sp})
certainly give the maximum of $\text{Re}\{G(y)\}$.
The contributions from these $y_k$'s can be summed up explicitly,
and the result is 
\begin{align}
  \label{spt}
  \tilde{Z}
     &\approx  2^N\sum_{k=-\i}^\i e^{NG(y_k)}\int_{-\i}^{\i}\frac{dy}{\pi}
            e^{\frac{N}{2}G''(y_k)y^2} \notag\\
     &=      \frac{2^N}{\beta\sqrt{\frac{\pi}{2} \sum_{j=1}^N{a_j}^2}}
            \sum_{k=-\i}^\i\frac{(-1)^{k\lambda}}{1+(\frac{\pi}{\beta} k)^2}
\nonumber\\
     &=
        \begin{cases}
          \frac{2^N}{\sqrt{\frac{\pi}{2}
\sum_{j=1}^N{a_j}^2}}\coth\beta 
          & (\lambda:\text{even})\\
           \frac{2^N}{\sqrt{\frac{\pi}{2} \sum_{j=1}^N{a_j}^2}}
          \text{cosech}\beta  
          & (\lambda:\text{odd})
        \end{cases},
\end{align}
where $\lambda=\sum_{j=1}^{N} a_j$.
Here we remark that in \cite{mPRL98,mP00} the formula for the case of
odd $\lambda$ is missing.
(Also, in the arguments of the constrained case in \cite{mP00},
the results are presented only for the case of even $N$, but
one has to consider the case of odd $N$ separately.)
Nevertheless, we only consider the case where $\lambda$ is even 
in the following, because the odd $\lambda$ case can be discussed 
similarly.

One notes that the expression (\ref{spt})
diverges as $\beta \to 0$ while the correct limiting value is clearly
$2^N$ from the definition (\ref{MZ}).
Hence, as mentioned in the main text, his result (\ref{spt}) is 
not valid at least for small $\b$ (or large $T$).

In \cite{mPRL98,mP00}, 
the hard region was also discussed using (\ref{spt}).
In particular, the average minimum cost was estimated.
One should use the finite-temperature expression of the partition 
function (\ref{spt}) to analyze the non-vanishing value of 
average minimum cost in the hard region. However, since (\ref{spt})
is not reliable for large $T$, the formulas given in \cite{mPRL98,mP00}
should be taken with special caution.
The principal source of trouble is  in the anomalous properties
of the systems with the decreasing number of configurations
as the energy increases. There is another point of problems in 
his analysis as we discuss in the following.

A sign of difficulty 
is seen from the negative value of the entropy in the hard region.
The entropy calculated from the partition function (\ref{spt}) reads
\begin{equation}
\tilde{S}=\log \frac{2^N}{\sqrt{\frac{\pi}{2} \sum_{j=1}^N
{a_j}^2}}\coth\beta+ \frac{\beta}{\sinh
\beta \cdot \cosh \beta}.
\end{equation}
The ground state entropy $\tilde{S}_0=\lim_{\b\to\i} \tilde{S}$ is found to be
\begin{equation}
\tilde{S}_0 = \{N-N_c({\mathcal A})\}\log 2, \label{eqn:3-S}
\end{equation}
with
\begin{equation}
N_c({\mathcal A})=\frac{1}{2}\log_2\frac{\pi}{2} \sum_{j=1}^N {a_j}^2.
\end{equation}
This is equivalent to (\ref{W_b0}).
One notices that, when $N < N_c({\mathcal A})$, 
the ground state entropy is negative.
In \cite{mPRL98,mP00}, the easy (resp. hard) region is characterized by 
a positive $\tilde{S}_0$ (resp. a negative $\tilde{S}_0$), i.e., 
by $N>N_c(\mathcal{A})$  (resp. $N<N_c(\mathcal{A})$). 
In an appropriate limit, this coincides with $\k<\k_c$ (resp. $\k >\k_c$).
To avoid the difficulty of negative entropy in the hard region, 
the author of \cite{mPRL98,mP00} proposed not to take the 
$\b\to\i$ limit but to use (\ref{spt}) only down to the 
temperature where $\tilde{S}\geq \log 2$. This is an arbitrary process
which would not be necessary if we use the exact expression of the entropy.

To identify the problem within his formalism, let us
remember that (\ref{spt}) was obtained by summing up only
the contributions from around extreme points $\{y_k\}$ of (\ref{3-sp}). 
In the easy region $(N \gg N_c({\mathcal A}))$ the peaks around these
points are very sharp
and hence the Laplace method gives a good approximation.
On the other hand, in the hard region $(N \ll N_c({\mathcal A}))$, 
there appear a large number of other local maxima,
with values not so far from zero and 
at points located fairly close to the points  $\{y_k\}$ 
of (\ref{3-sp}).
One will be easily convinced that this happens by checking
a very simple example of $N=2$.
In this case $G(y)$ reads
\begin{equation}
\label{N2}
G(y)= \frac{1}{2}\log\cos(\beta a_1\tan y)
      +\frac{1}{2}\log\cos(\beta a_2\tan y),
\end{equation}
where $a_1,a_2(a_1 < a_2)$ are coprime natural numbers with $a_2$
sufficiently large corresponding to the hard region.

\newpage
\begin{large}
\noindent
Figure Captions
\end{large}

\vspace{10mm}
\noindent
Fig. 1: 
The number of solutions $W(E)$ as a function of the energy $E$
for an example with $N=20,L=256$, and 
$\mathcal{A}=
\{218,13,227,193,70,134,89,198,205,147,227,190,27,239,192,131\}$.
The theoretical prediction is indistinguishable from the numerical 
results plotted in dots.

\vspace{10mm}
\noindent
Fig. 2: 
The easy/hard regions of the randomized subset sum.


\renewcommand{\thepage}{Figure 1}
\begin{picture}(400,300)
\psfrag{WLER}{\hspace*{-3mm}\large $W(E)$}
\psfrag{E}{\large $E$}
{\includegraphics{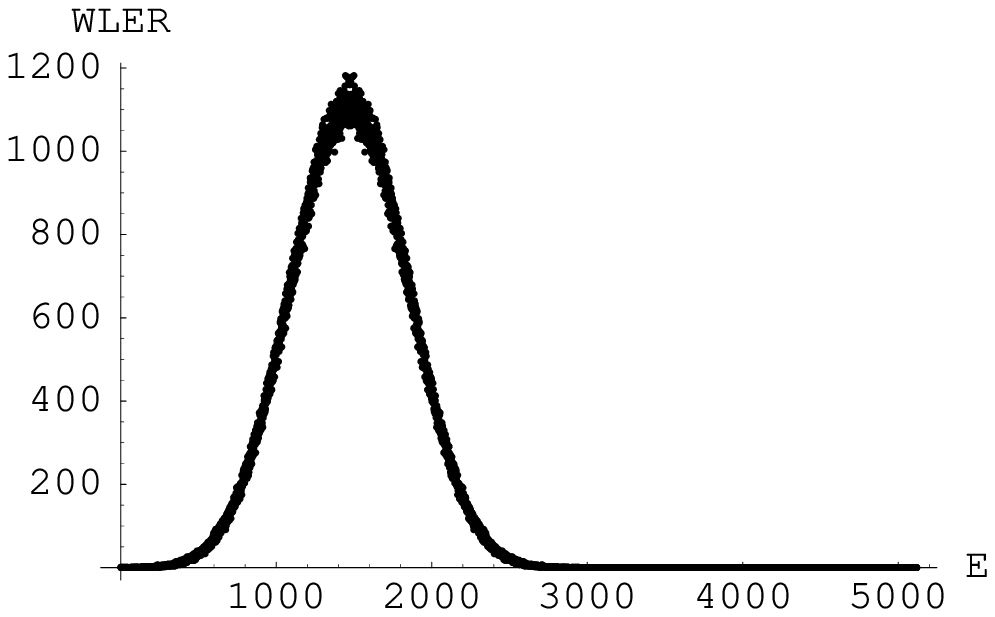}}
\end{picture}

\newpage
\renewcommand{\thepage}{Figure 2}
\begin{picture}(400,300)
\put(90,60){\Large easy}
\put(210,110){\Large hard}
\psfrag{x}{\LARGE $x$}
\psfrag{k}{\LARGE $\k$}
{\includegraphics{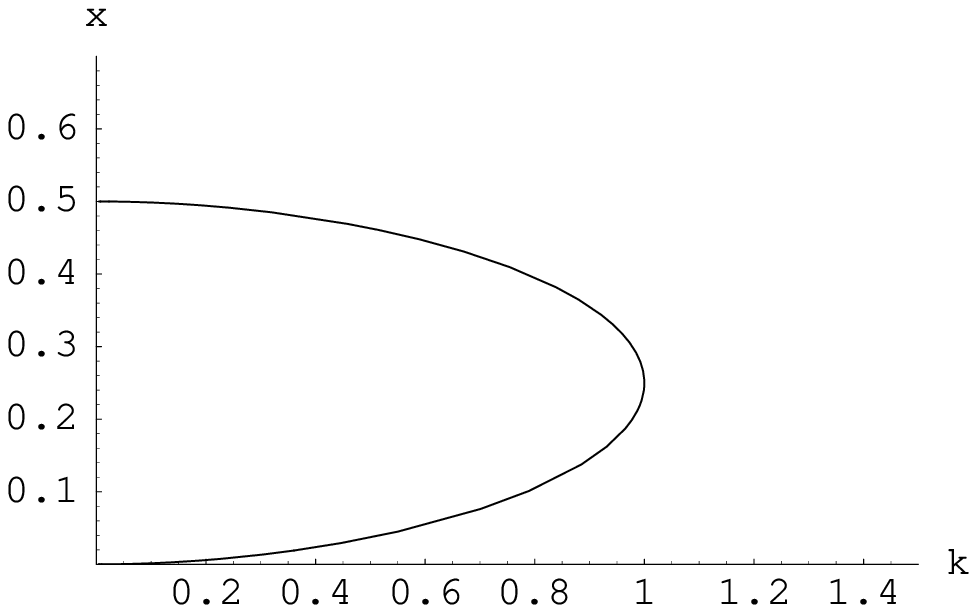}}
\end{picture}

\end{document}